\documentclass[11pt,twoside]{article}

\usepackage{asp2004}
\usepackage{epsf}
\usepackage{psfig}
\usepackage{lscape}

\begin{document}
\title{Gaia --- A White Dwarf Discovery Machine}   %%% Fill in title
\author{Stefan Jordan}   %%% Fill in author names
\affil{Astronomisches Rechen-Institut, Zentrum f\"ur Astronomie der Universit\"at Heidelberg, M\"onchhofstr. 12-14,
       D-69120 Heidelberg, Germany}    %%% Fill in author affiliations

\begin{abstract} %%% Abstract to run on from here.
Gaia is a satellite mission of the ESA, aiming at absolute astrometric measurements of about one billion stars
($V<20$) with unprecedented accuracy. Additionally, magnitudes and colors will be obtained for all these stars, while
radial-velocities will be determined only for  bright objects ($V<17.5$). However, the wavelength range for the
radial-velocity instrument is rather unsuitable for most white dwarfs.
Gaia will probably discover about 
400,000 white dwarfs; up to 100\,pc the detection probability for white dwarfs is almost 100\,\%. 
This survey of white dwarfs will have very clear, easy to understand selection criteria, and will therefore
be very suitable for statistical investigations. The Gaia data will help to improve
the construction of a luminosity function for the disk and the halo 
and will provide a more accurate determination of the age of our solar neighborhood. 
Moreover, reliable stellar dynamical investigations of the disk and halo components will be possible. 
For the first time it  will be possible to test the mass-radius relation of white dwarfs in great detail.
Moreover, more accurate masses of magnetic and cool white dwarfs can be expected. Gaia is also expected to discover
many new pulsating white dwarfs.
The Gaia measurements can also complement the measurements of gravitational waves from close white dwarf binaries with
Lisa. 
\end{abstract}

%%% MAIN BODY OF TEXT GOES HERE. CONSULT "INSTRUCTIONS FOR AUTHORS USING
%%% LATEX2E MARKUP", SECTIONS 2.3-2.6 FOR HELP WITH EQUATIONS, FIGURES,
%%% AND TABLES.

\section{The Gaia mission}
The orbit of the Gaia mission has been chosen to be a controlled Lissajous orbit around the Lagrangian point L2 of the Sun-Earth system
in order to have a quiet environment for the payload
in terms of mechanical  and  thermo-mechanical stability. Another advantage of this position is
the possibility of uninterrupted observations, since the Earth, Moon and Sun all lay within Gaia's orbit.
The aim of Gaia is to perform absolute astrometry rather than differential measurements in a small field of view.
For this reason, Gaia -- like Hipparcos -- (i) simultaneously observes in two fields of view (FoVs) separated by a large
basic angle of 106.5$^\circ$, (ii) roughly scans along a great circle leading to strong mathematical closure conditions,
(iii) performing mainly one-dimensional measurements, and (iv) scanning the same area of sky many times during the mission
under varying orientations. 

\begin{figure}[t]
\plotfiddle{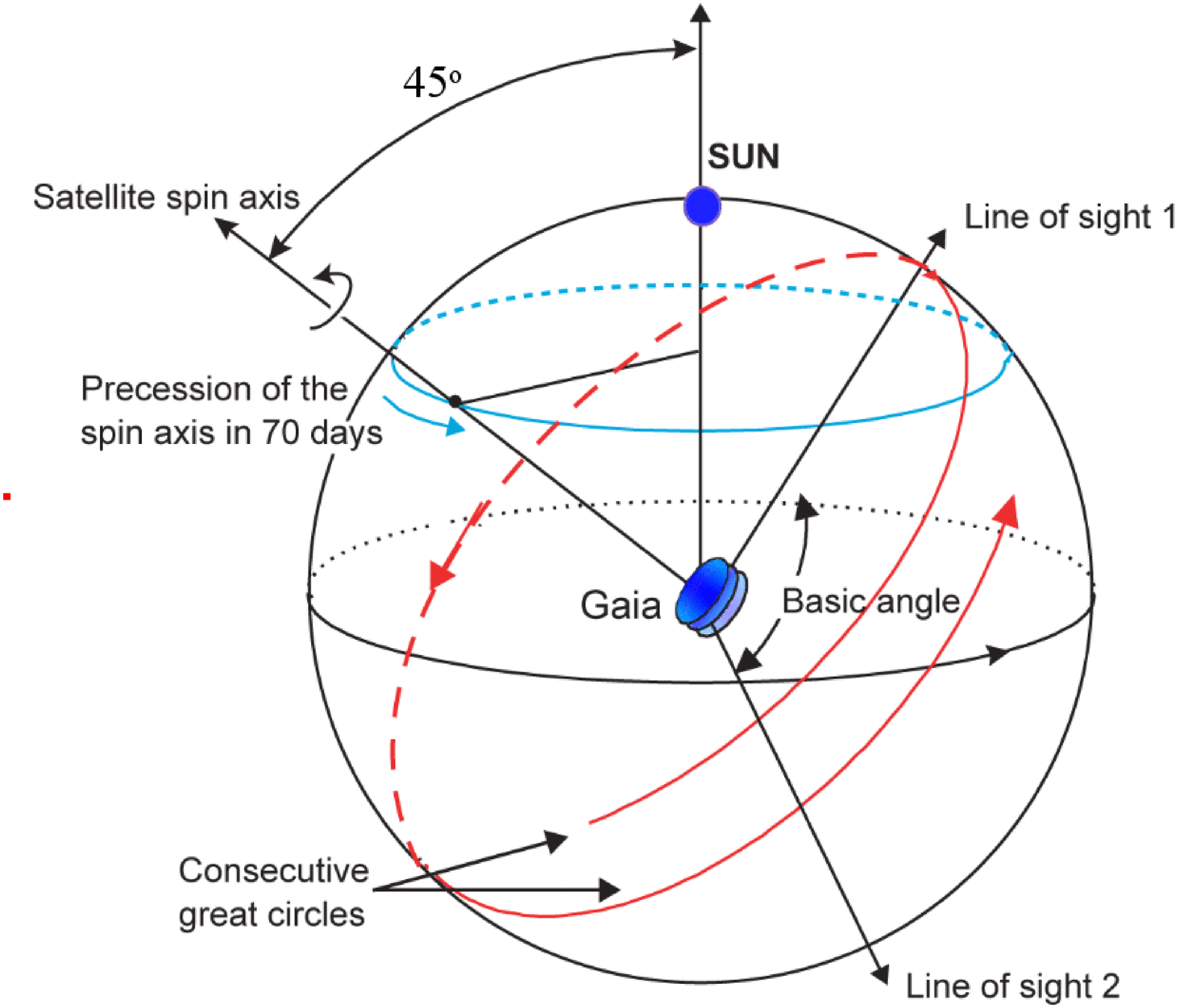}{190pt}{0}{25}{25}{-100}{10}
\caption{Gaia's two fields of view, separated by a basic angle of 106.5$^\circ$  scan the sky according to a
nominal scanning law: Rotation period: 6 hours, solar aspect angle: 45$^\circ$, precession period: 63~days.
\label{nsl}
}
\end{figure}

These conditions are fulfilled by Gaia's nominal scanning law (see Fig.\,\ref{nsl}):
The satellite will spin around its axis with a constant rotational period of 6 hours. The spin axis will precess
around the solar direction with a fixed aspect angle of 45$^\circ$ in 63.12 days. On average, each object in the sky is transiting the
focal plane about 70 times  during the 5 year nominal mission duration. Most of the times, an object transiting through
one FoV is measured again after 106.5 or 253.5 minutes (according to the basic angle of 106.5$^\circ$) in the 
other FoV.

The Gaia payload consists of three instruments mounted on a single optical bench: The astrometric instrument,
the photometers, and a spectrograph to measure radial velocities.

\begin{figure}[t]
\plotone{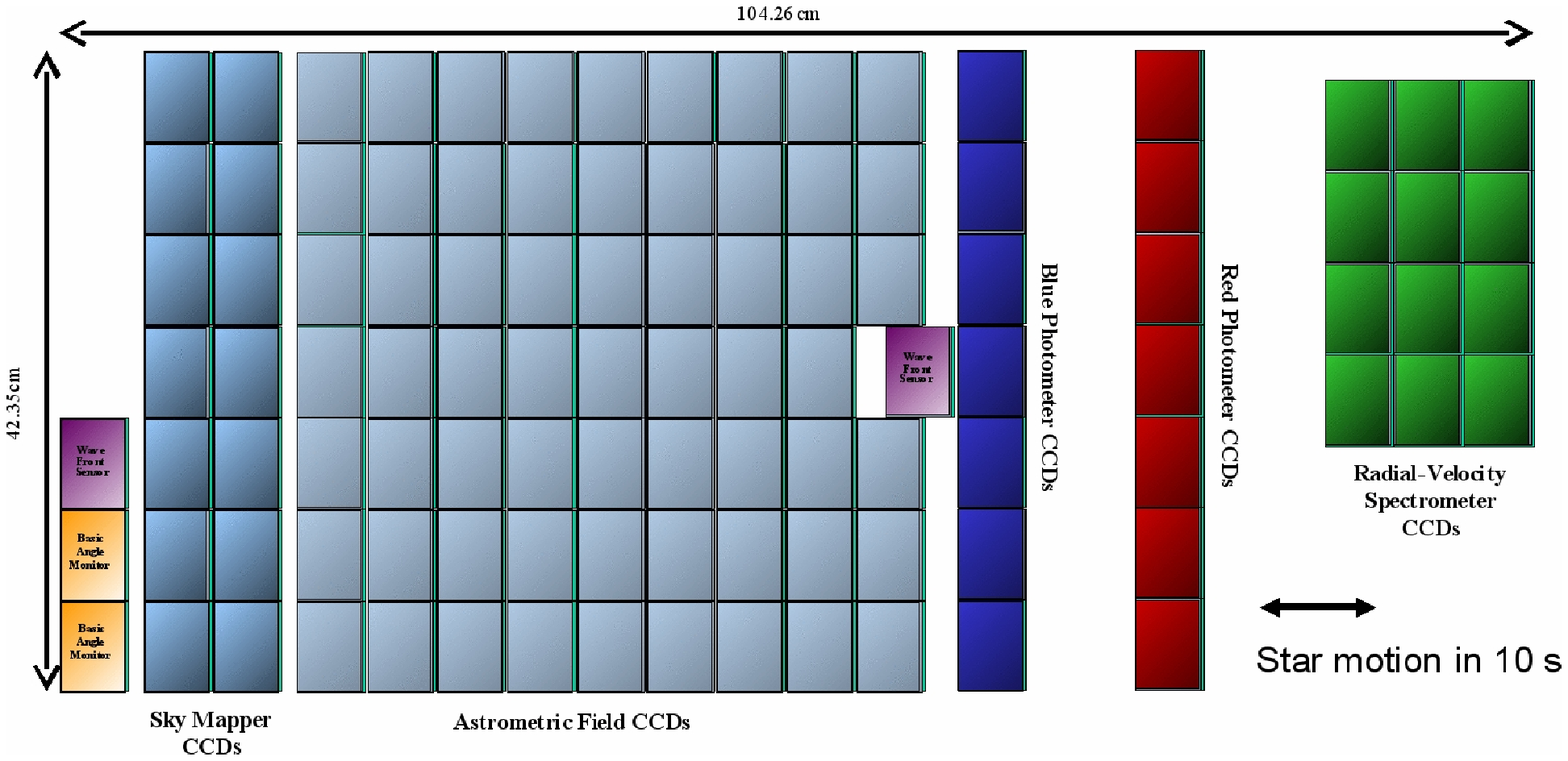}
\caption{Focal Plane Assembly. Each CCD has a size of 
  about  4\,cm$\times$6\,cm. The direction of star image motion is
  indicated at the bottom. It takes a star image about 4.4\,sec to cross
  one of the CCDs. 
  Instruments: The sky mappers
  (SM), the main astrometric field (AF), the blue photometer (BP), the
  red photometer (RP), the radial-velocity spectrograph (RVS),
  the basic
  angle monitoring (BAM) system, and the wavefront sensors (WFS).
\label{FPA}
}
\end{figure}

The astrometric field consists of 62 CCDs and a star is measured on 8-9 CCDs during one transit (see Fig.\,\ref{FPA}).
 The accumulated charges
of the CCD are transported across the CCD in time delay integration mode in synchrony with the images.
 In order to reduce the data rate and the read-out noise only small windows
around each target star, additionally binned in across-scan direction depending on the object's magnitude,
are read out and transmitted to the ground.

Multi-colour photometry is provided by 
two low-resolution fused-silica prisms dispersing all the
light entering the field of view in the along-scan direction prior to detection.
The Blue Photometer (BP) operates in the wavelength
range 3300--6800\,\AA; the Red Photometer (RP) covers the
wavelength range 6400--10500\,\AA.  

The RVS is a near infrared (8470 -- 8740\,\AA), medium resolution spectrograph:
$\mathrm{R} = \lambda / \Delta \lambda = 11\,500$. It is illuminated by the same
two telescopes as the astrometric and photometric instruments.  

The  astrometric core solution will be based on about  $10^8$ primary stars which means to solve for
some $5\times 10^8$ astrometric parameters (positions, proper motions, and parallaxes). However, the attitude of the
satellite (parameterized into $\sim 10^8$ attitude parameters over five years) can also only be determined with high 
accuracy from the measurements itself. Additionally, a few million calibrational parameters describe the geometry of the
instruments, and finally, deviations from general relativity are accounted for by solving for the
post-Newtonian parameter $\gamma$. 

The number of observations for the $10^8$ primary stars is about $6\cdot 10^{10}$.
The condition equations
connecting the unknowns to the observed data are 
non-linear but linearize well at the
sub-arcsec level. Direct solution of the corresponding least-squares
problem is infeasible, because 
the large number of unknowns and their strong inter-connectivity,
prevents any useful decomposition of the problem into
manageable parts. The proposed method is a block-iterative scheme 
called the Global Iterative Solution. Intensive tests are currently under way and  have already
demonstrated it's feasibility with $10^6$ stars assuming realistic random and systematic 
errors in the initial conditions.

Later, with a good solution for the attitude and the geometric parameters based on the 
measurements of the  $10^8$ primary stars, the remaining $9\cdot 10^8$ stars can
be linked into the system. 

The precision of the astrometric parameters of individual stars
depends on their magnitude and color, and to a lesser
extent on their location in the sky. Sky-averaged values
for the expected parallax precision are displayed in
Table\,\ref{tab:astrom-performance}. The corresponding figures for the
coordinates and for
the annual proper motions are similar but slightly smaller
(by about 15 and 25\%).
Note, that a parallax accuracy of  25 microarcseconds (the thickness of a human hair seen from a distance of 200 km)
 means an accuracy of the distance determination of 
0.1\%\ for 40\,pc, and 1\%\ for 400\,pc.

Additionally to the astrometric measurements the entire sky will be observed with the same spectro-photometric system and with
unprecedented homogeneity. Moreover, radial velocities are measured with a precision between 1\,km/sec (V=11.5) and
30\,km/sec (V=17.5). Photometric measurements are not only nice to have but will allow the correction of 
relative displacement between an
early-type and  very red stars (chromaticity) which may be as large as 1 milliarcsecond.
The measurements of radial velocities are important to correct for perspective acceleration which is induced by the
motion along the line of sight.

\begin{table}[htb]
\caption{End-of-mission parallax precision in microarcseconds.
  Representative values are shown for unreddened stars of the indicated
  spectral types and V~magnitudes.
  The values are computed 
  using the actual Gaia design as input.
  The performance calculation used does not include the effects of radiation
  damage to the CCDs. 
  \label{tab:astrom-performance}}
\begin{center}
\begin{tabular}{|cc|r|}
\hline
Star type & V magnitude   & 2006 nominal \\
          &               & performance  \\
\hline
          &  $<$ 10     &   5.2        \\
B1V       &   15        &   20.6       \\
          &   20        &  262.9       \\
\hline
          &   $<$ 10    &   5.1        \\
G2V       &   15        &   19.4       \\
          &   20        &  243.4       \\
\hline
         &  $<$ 10     &   5.2        \\
M6V       &   15       &   8.1        \\
          &   20       &   83.9       \\
\hline
\end{tabular}
\end{center}
\end{table}

Gaia is currently scheduled to be launched from Kourou, French Guiana,
in December  2011 with a Soyuz-ST rocket (which includes a restartable Fregat upper stage). Initially 
the Fregat-Gaia composite is placed into a parking orbit, after which a single Fregat boost injects Gaia
on its transfer trajectory towards the L2 Lagrange point. In order to keep Gaia in an orbit around L2, the spacecraft must
perform small maneuvers every month. 
After a commissioning phase Gaia will measure the sky for five years with a possible extension for
another year. Subsequently, the final catalog, which includes astrometric and photometric information,
radial-velocity determinations, and a classification of the objects, will be produced. The completion of the Gaia mission
is intended to be around 2020 (at the time of the 22$^{\rm th}$ European Workshop on White Dwarfs?).
 However, it is not implausible that preliminary products will be delivered earlier, since
even the results from only one year of measurements would considerably surpass the precision of all existing 
star catalogs.

\section{Gaia's performance for white dwarfs}   
The number of currently known white dwarfs amounts to almost 10,000, 5500 in the online version of the Villanova White Dwarf Catalog \footnote{http://www.astronomy.villanova.edu/WDCatalog/index.html}, 9316 confirmed white dwarfs in the SDSS according to 
\cite{Kleinman-et-al:eurowd}.
\cite{Torres-et-al:05} have performed intensive Monte-Carlo simulations and arrived at about 400,000 white dwarfs
down to $V=20$ that will be detected by Gaia. For disk white dwarfs Gaia will be practically complete up to
100\,pc and will observe about half  of all white dwarfs within  300\,pc, decreasing to one third at distances of
400\,pc.  Disk white dwarfs at the cut-off of the luminosity function ($M_{\rm bol}\approx 15.3$, $M_V\approx 16$) can be
detected up to distances of 100\,pc, considerably improving the age determination of the solar neighborhood to about
$\pm 0.3$\,Gyr. Moreover, a detailed check of white dwarf cooling theory is possible by a careful analysis of the
Gaia white dwarf luminosity function.

The question, whether Gaia can distinguish between disk and halo stars was also investigated by \cite{Torres-et-al:05}.
Since the wavelength range of the Gaia radial-velocity spectrograph (8470 -- 8740\,\AA) does not allow the measurement
of radial velocities for almost all white dwarfs, they used only a reduced-proper motion criterion which additionally
throws away the direction of the proper motion. From this they concluded that the distinction between both populations is
extremely difficult. However, if one uses the direction information, halo white dwarf can be detected more easily, as
was e.g. demonstrated by \cite{Carollo-et-al:eurowd}. With the Gaia sample, a detailed investigation of 
different disk components (``thin disk'', ``thick disk'') and their scale heights will also be possible. Moreover,
distinctive luminosity functions can be constructed for all disk and halo components, considerably improving our
understanding of our Galaxy's formation.

Due to the very large errors in the distance determination even one of the fundaments of the theory of white 
dwarfs is not yet tested with sufficient accuracy: the mass-radius relation. \cite{Vauclair-et-al:97} used 22
state-of-the-art ground-based or Hipparcos parallaxes (with 13\%\ error on average) and were unable to 
account for the theoretical mass-radius relation. \cite{Provencal-et-al:02} used only the best three white dwarf
parallaxes measured by Hipparcos  (3\%\ error on average) and came somewhat closer to the goal of verifying  the theory.
The number of white dwarfs with high-precision distance determinations is simply too small for a detailed analysis
of this question.
With the typical 0.1\%-1\%\ accuracy of the Gaia parallaxes, it will be possible to study the dependence of the
mass-radius relation on $T_{\rm eff}$ and the chemistry in great detail, with the further ability to discriminate between
different hydrogen envelope masses. 

Moreover, by such an investigation Gaia will certainly find out whether strange things like strange matter 
\citep{Fontaine-et-al:eurowd} or strange interiors like iron cores \citep{Shipman-Provencal:99} exist!

Gaia will also detect a large number of spectroscopic white dwarf binaries by finding strong disagreement between spectroscopic
and astrometric parallaxes. The discovery of close spectroscopic binaries is very important for the identification of sources
for gravitational waves measured by the Lisa mission. Independent measurements of masses and separations from the 
Lisa data are only
possible with accurate distances \citep{Stroeer-et-al:05}.

In the case of strongly magnetic white dwarf, the simultaneous presence of the Zeeman and Stark effect does not allow
a reliable estimation of $\log g$. Therefore, masses can only be  quantified by combining photometry, spectroscopic
determinations of effective temperatures, and parallaxes. Therefore, this field will also considerably benefit  from
the Gaia measurements, and the question whether magnetic white dwarfs are  more massive  \citep{Liebert:88} for all
field strengths 
can be scrutinized together whith it's consequence for the identification of the main-sequence progenitors.

However, the inability to determine reliable masses is not only limited to magnetic white dwarfs. As was 
disscussed in detail \cite{Bergeron-et-al:eurowd} and \cite{Kepler-et-al:eurowd}, the masses of cool white dwarfs
($T_{\rm eff}<10000$\,K),  as inferred from spectroscopy,  strongly deviate from those at higher temperatures.
With the addition of reliable distance determinations this mystery can certainly be solved.
The question, whether DAs and non-DAs have the same mass distribution can also be answered with a higher precision.

Not only the astrometric measurements will be important for white dwarf research: Many new non-radial pulsators
will be found by analyses of the high-precision photometry. The nominal scanning law will lead
to a very non-uniform sequence  of detections in time: one FoV transit corresponds to 1.5 minutes, 
the time separation between the two FoVs is 1.8 or 4.2 hours, and up to 14 transits within about 40\,hours 
are possible; smaller groups of observations are more frequent, but most groups contain at least two transits.
The question, which periods and amplitudes are detectable has to be studied in more detail, but it is certain
that the number of confirmed pulsating white dwarfs will increase significantly.

Additionally to all these details, one of the major advantages of the Gaia sample of  white dwarfs
is that it constitutes  an all-sky survey with very clear selection criteria. Biases introduced by 
unclear selection effects are certainly one of the major obstacles of statistical investigations of white
dwarfs \citep{Kleinman-et-al:eurowd}.

One can clearly conclude that white dwarf research will tremendously benefit from the Gaia data.
This is not surprising, because there is hardly any topic in astrophysics that will be not affected
(at least indirectly) by the measurements of this satellite mission!

\acknowledgements 
Work on Gaia data processing in Heidelberg  is supported by the
DLR grant 50 QG 0501. The figures are courtesy of EADS Astrium and ESA.

%%% THE BIBLIOGRAPHY
%%%
%%% CONSULT SECTION 3 OF "INSTRUCTIONS FOR AUTHORS" FOR HOW TO USE NATBIB.
%%% AUTHORS ARE ENCOURAGED TO USE EITHER THE "THEBIBLIOGRAPY" ENVIRONMENT
%%% BY UNCOMMENTING (DELETING THE "%" SYMBOL) THE COMMANDS BELOW, OR BY
%%% USING THE BIBTEX ENVIRONMENT. TO FIND OUT WHICH IS APPLICABLE TO YOUR
%%% CONTRIBUTION, CONSULT THE VOLUME EDITORS FOR YOUR PROCEEDINGS.
%%%


\begin{thebibliography}{}
\bibitem[{Bergeron et al. (these proceedings)}]{Bergeron-et-al:eurowd}
       Bergeron, P., Gianninas, A., \& Boudreault, S., these proceedings
\bibitem[{Carollo et al. (these proceedings)}]{Carollo-et-al:eurowd}
       Carollo, D., Bucciarelli, B., Hodgkin, S. T., et al., these proceedings
\bibitem[{Fontaine et al. (these proceedings)}]{Fontaine-et-al:eurowd}
       Fontaine, G., Bergeron, P., \& Brassard, P.,  these proceedings

\bibitem[{Kepler et al. (these proceedings)}]{Kepler-et-al:eurowd}
       Kepler, S. O., Kleinman, S. J., Nitta, A., Koester, D., Castanheira, B. 
G., Giovannini, O., Althaus, L.,  these proceedings
\bibitem[{Kleinman et al. (these proceedings)}]{Kleinman-et-al:eurowd}
       Kleinman, S. J., Eisenstein, D. J., Liebert, J., \& Harris, H. C., these proceedings
\bibitem[{Liebert (1988)}]{Liebert:88}
         Liebert J., 1988, PASP 100, 1302
\bibitem[{Provencal et al. (2002)}]{Provencal-et-al:02}
Provencal J. L., Shipman H. L., Koester D., Wesemael F., Bergeron P., ApJ 568, 324
\bibitem[{Shipman \&\ Provencal (1999)}]{Shipman-Provencal:99}
 Shipman H. L., \&\ Provencal J. L. 1999, in ASP Conf. Ser. 169, 11th European Workshop on White Dwarfs, ed. J.-E. Solheim \&\ E. G. Mei{\^s}tas (San Francisco: ASP), p. 293
\bibitem[{Stroeer et al.  (2005)}]{Stroeer-et-al:05}
         Stroeer A., Vecchio A., Nelemans G., ApJ 633, L33
\bibitem[{Torres et al. (1999)}]{Torres-et-al:05}
         Torres S., Garc{\'i}a-Berro E., Isern J., Figueras F.,
          MNRAS 360, 1381
\bibitem[{Vauclair et al. (1997)}]{Vauclair-et-al:97}
        Vauclair G., Schmidt H., Koester D., Allard N., A\&A 325, 1055
\end{thebibliography}
\end{document}